\documentstyle[12pt]{article}
\def\beq{\begin{equation}}
\def\eeq{\end{equation}}
\def\beqar{\begin{eqnarray}}
\def\eeqar{\end{eqnarray}}

\def\he#1{\hbox{${}^{#1}$He}}
\def\li#1{\hbox{${}^{#1}$Li}}

\def\yp{\hbox{$Y_{\rm p}$}}

\def\la{\mathrel{\mathpalette\fun <}}

\def\fun#1#2{\lower3.6pt\vbox{\baselineskip0pt\lineskip.9pt
  \ialign{$\mathsurround=0pt#1\hfil##\hfil$\crcr#2\crcr\sim\crcr}}}

\begin{document}

\begin{titlepage}
\pagestyle{empty}
\baselineskip=21pt
\rightline{UMN-TH-1407/95}
\rightline{hep-ph/yymmddd}
\rightline{August 1995}
\vskip .2in
\begin{center}

{\bf MODEL-INDEPENDENT PREDICTIONS OF \\
BIG BANG NUCLEOSYNTHESIS}

\bigskip

Brian D. Fields$^{1,2}$ and  Keith A. Olive$^3$

\bigskip

{\it
$^1$Institut d'Astrophysique, 98 bis boulevard Arago, Paris 75014 France \\
\medskip
$^2$Department of Physics, University of Notre Dame, Notre Dame, IN 46556 \\
\medskip
$^3$School of Physics and Astronomy, University of Minnesota,
Minneapolis, MN \\}
\medskip

\bigskip
\bigskip

{\bf Abstract}

\medskip

\end{center}

 We derive constraints  from standard (with $N_\nu = 3$)
BBN arising solely from two cosmologically produced nuclides,
\he4 and \li7, from which the extrapolation to primordial
abundances is straightforward.  The abundances of D and \he3
are at present only inferred from their solar and local
interstellar medium values using models of galactic
chemical evolution. However, our knowledge of chemical
evolution suffers from large uncertainties,
and so it is of use to take an approach which minimizes the
reliance on chemical evolution in determining the consistency of
BBN. Using only the data on \he4 and \li7, and reasonable estimates
of possible systematic errors in their abundance determinations,
concordance is found if the baryon-to-photon ratio lies in the
range $1.4 < 10^{10} \eta < 3.8$ (95\% CL). The \he4 and \li7 abundances
are also used to predict  the primordial abundances
D and \he3, which provides an initial condition for their chemical
evolution.

\noindent
\end{titlepage}
\baselineskip=18pt

The consistency of big bang nucleosynthesis (hereafter BBN) is
of major importance to standard cosmology, as BBN tests
the big bang at the earliest epoch accessible thus far
 \cite{wssok}.
While BBN theory is well-understood,
determining its consistency with the data
is difficult because the observables are not the primordial abundances
themselves, but rather present-day or solar abundances.
We have, at present, only local
and recent
measurements from which we must infer
the primordial light element abundances.
Such an extrapolation falls within
the domain of chemical evolution, a discipline
fraught with its own difficulties and uncertainties.  In this paper, we
will seek to minimize the use of chemical evolution in determining
the viability of BBN, and indeed we will turn the problem around and
offer BBN results as constraints on chemical evolution models.

For clarity, we explicitly distinguish the
three features inherent to ``BBN'' and required in evaluating
its consistency.  (1) Theory, i.e., the calculation of light
element abundances as a function of the baryon-to-photon ratio
$n_{\rm B}/n_\gamma \equiv \eta$.
We will consider here
only the standard model of big bang nucleosynthesis with
three massless neutrinos, $N_\nu = 3$; with $N_\nu$ fixed,
$\eta$ is the only free parameter.
The calculation  proceeds
from first principles and is well understood;
its uncertainties are those due to nuclear cross sections
measured in the lab and so admit the usual error analysis techniques.
This has led to monte carlo calculations of the big bang abundances
\cite{kr} - \cite{hata1},
all of which are in good agreement.
(2) Data, i.e.,  the observed light element
abundances in stars, or in galactic and
extragalactic gas.  The observations have
varying degrees of statistical uncertainty, but
the more serious problem is that the systematic uncertainties,
while hard to determine, could
well dominate the error in the data.  (3) Chemical evolution,
which accounts for galactic nucleosynthesis processing in the observed
abundances and so allows extrapolation
to the primordial abundances needed to test the theory.
Beyond a set of general principles, chemical evolution is
very uncertain in its details, and we are lacking as yet a first-principles
theory, relying rather on a simplified (yet in some aspects,
 surprisingly successful)
model, which itself has many uncertain parameters
(see e.g.  \cite{bt}).
Notice that BBN is an empirically testable theory only with all
three features, in particular, only with the introduction, at some
level, of chemical evolution.

Given that the detailed predictions of chemical evolution
have questionable reliability and accuracy, and also given
the need to include chemical evolution in the evaluation of BBN,
several approaches suggest themselves.
One is to try to find the most generic (and hopefully model-independent)
predictions of chemical
evolution, either by using the best available parameterizations
and examining the range in which they may vary \cite{st92}
 - \cite{fields},
or by making a simple parameterization of the basic predictions
as relevant to BBN \cite{st95} - \cite{cstb}.
Such an approach makes all four light element species available to
test BBN, and so can provide a powerful test of the theory as the
(one-parameter) predictions are strongly over-constrained.
However, such an approach must always be subject to criticisms
of the chemical evolution procedure used, particularly for elements
for which the effects of evolution are large.

Another means of addressing the role of chemical evolution uncertainties
in evaluating BBN---and the approach we will take---is to see what
conclusions one can reach with
a minimal reliance on chemical evolution.  To do so, we first note that
the different light nuclides have very different degrees of galactic
evolution.  Specifically, to infer
the primordial abundances of \he4 and \li7 involves only an
appeal to the most general principles of chemical evolution (namely
the increase of metal abundances with time) while to determine
D and \he3 evolution relies on calculations that are much more detailed
(and thus much more model-dependent).
Consequently, for the purposes of evaluating BBN,
we will restrict ourselves to just the abundances of \he4 and \li7; note that
these two abundances alone are enough to test the one-parameter,
standard BBN theory.  We will thus see if there
is a concordant range in $\eta$ for current best estimates of
primordial \he4 and \li7, and indeed we find one.  Then we
set confidence limits on BBN predictions for primordial D and \he3, which
then serve as inputs for chemical evolution models.

Our approach to obtaining bounds on $\eta$ is clearly different from previous
ones.  In particular, the lower bound on $\eta$ was first established using
the combination of D + \he3 in \cite{ytsso}.  This argument,
 which is based on the fact \he3 is not totally destroyed in stars, was used
and/or improved upon in later work \cite{wssok,hata2,cstb}. It has been
argued \cite{hata2} that the lower bound on $\eta$ is now large enough
to indicate a potential inconsistency in BBN. Indeed, when \he3 production in
low mass stars is included \cite{orstv}, (this inclusion is implied by recent
observations of high \he3 in planetary nebulae \cite{rood})
one is faced with an overproduction of \he3.  As we state below, we will
not make any speculations on the source of the problem concerning \he3 here
(see e.g. \cite{scostv}).
It is clear however, that the assumed primordial abundances of D and \he3
will be strongly dependent on models of chemical and stellar evolution.
It is for this reason, that we will perform our analysis independently
  of these two isotopes. Due to our analysis based only on \he4 and \li7, our
conclusions regarding the consistency range for $\eta$ are also different
than in previous work.  We will find that a lower value for $\eta$ is preferred
and that overlap with other work occurs at about the $2\sigma$ level.
In addition, our predicted range for primordial D/H is correspondingly
higher than in previous work.

This straightforward analysis is complicated by the need to consider
the presence of significant systematic errors in the \he4 and \li7
abundances.  While such errors are intrinsically hard to quantify,
they are likely to be present and could in fact dominate the error.
We will thus consider several possibilities for the size and distribution
of the systematic errors, and their possible combinations for the
two elements.

\bigskip

The light elements are observed in disparate astrophysical sites,
from old stars to galactic as well as
 extragalctic gas.  Questions of chemical evolution
aside, the abundances themselves are difficult to determine to the
needed accuracy.  Systematic errors can arise in the procedure
used to deduce an abundance from a line strength, and from the
idealizations employed in modeling the sites themselves.
In the following we briefly review the situation for each
light nuclide, with attention to the possible systematic errors.

\bigskip
\noindent{\it \he4:}
\bigskip

The \he4 abundance  is best determined
from observations  of HeII $\rightarrow$ HeI
recombination lines in extragalactic HII regions.
There are extensive compilations of observed
abundances of \he4, as well as the abundances of N, and O,
in many different low metallicity HII regions
\cite{p}- \cite{iz}.
The oxygen abundance in these regions ranges from one fifth to one
fiftieth of the solar oxygen abundance. However because \he4 is
produced in stars along with oxygen, the primordial abundance of
\he4 can only be determined from an extrapolation of the data to
zero metallicity. Fortunately there is data at low metallicity which
lends confidence to such an extrapolation without the reliance of
specific models of galactic chemical evolution other than the assumption that
both oxygen (as well as nitrogen) and \he4 increase with time due to their
production in stars.  In an extensive analysis,
using the data of Pagel \cite{p} and Skillman  et al. \cite{evan},
Olive and Steigman \cite{osa} derived a primordial abundance of
$Y_p = 0.232 \pm 0.003$. With the inclusion of the recent
  data of Izatov et al.
\cite{iz}, the best estimate for the primordial \he4 mass fraction
becomes \cite{osc}
\beq
Y_p = 0.234 \pm 0.003 \pm 0.005
\label{eq:yp}
\eeq
where the first error in eq.\ (\ref{eq:yp}) is purely statistical.
The magnitude of the of the statistical uncertainty is dominated by the
large number of extragalactic HII regions observed (over 50) while
typical errors in any individual observation are of order 0.01.
One of the best observed HII region (and the one with the lowest metallicity),
I Zw 18, has an average \he4 abundance
\cite{p,evan2} of 0.229 $\pm$ 0.004.
There are in addition several sources of systematic uncertainties due
to ionization corrections, collisional excitation, and the presence of
neutral helium. The cumulative uncertainty from these effects has been
estimated to be of order 0.005 in the \he4 mass fraction
\cite{p,evan,osa}
though it may be somewhat higher \cite{sg,csta}.

\bigskip
\noindent{\it \li7:}
\bigskip

The primordial Li abundance is best determined from
observations of old, extremely metal poor (population II) halo stars.
These stars are observed to have a constant Li/H abundance
(the ``Spite plateau;'' \cite{ss})
below a metallicity less than about 1/20 of solar,
i.e., [Fe/H] $\la -1.3$.  Given that Fe must increase with
time in the Galaxy, the constant Li abundance for all low
metallicities indicates that the Li is primordial.
Notice that this conclusion relies on chemical evolution only
insofar as it is assumed that Fe increases with time in the early
Galaxy.

The pop II Li abundance is normally assumed to measure
 the primordial Li abundance;
and indeed Li has been measured in many such stars and so the
average abundance can in principle be determined to a high precision.
For a given method of converting the raw observations to an abundance,
the statistical errors are small. For an individual measurement, a \li7
abundance typically carries an uncertainty of 0.1 - 0.2 dex in [Li]
(\li7 abundances are normally quoted as logarithmic quantities so that
[Li] = 12 + $\log$ Li/H). Again, due to the large number of observations
the mean value can be determined to within 0.02 dex.
We will use the recent analysis of Molaro et al. \cite{mol}
in our computations
of BBN consistency. This leads to a mean value [Li] = 2.21 $\pm$ 0.02 or
\beq
\frac{\rm Li}{\rm H} = (1.6 \pm 0.1) \times 10^{-10}
\eeq

However, the overall accuracy of the observations suffers from two sources
of systematic error.  First, the
abundance determinations all depend on the model one adopts for
the stellar atmosphere; while different models (and different
researchers!) get roughly the same answer for the same stars,
some discrepancies still remain.
Thus we allow for a systematic error of magnitude
$\Delta \log_{10}({\rm Li/H}) = 0.10$ dex or $\Delta_1 = \Delta
 {\rm Li/H} = ^{+.4}_{-.3}
\times 10^{-10}$.
A second and potentially more serious problem is that the Li
may have been depleted over the long lifetimes of these stars,
and it
has been argued that rotational mixing could lead to very large depletions
\cite{ddp}.
While such models are hard to exclude, we note that the observations
(to include recent determinations of the \li7/\li6 ratio;
\cite{sln},
see also \cite{sfosw})
are well explained by non-rotational models \cite{ddk} which do not
give a significant depletion.  However, to be conservative, we will
examine the impact of allowing a Li depletion by a factor of 2.
Indeed it is also possible that some of the observed \li7 in halo
stars is not primordial and was produced by cosmic-ray nucleosynthesis.
While consistency of cosmic-ray nucleosynthesis with the observations
of Be and B in the same halo stars restricts the amount of cosmic-ray
produced lithium \cite{fossw},
 we can not be sure that some fraction of order 20\%
is not primordial. Thus we allow for a second systematic error in \li7
which we take as $\Delta_2 = \Delta {\rm Li/H} = ^{+1.6}_{-.3}
\times 10^{-10}$.

\bigskip
\noindent{\it D and \he3:}
\bigskip

We will not use D and \he3 to constrain BBN, but we will
compare BBN predictions
with the observed abundances to evaluate what chemical evolution models will
have to do.
The observational data is well reviewed elsewhere
(for a recent discussion see  \cite{scostv});
the upshot is that
solar abundances are
\beqar
\left( \frac{\rm D}{\rm H} \right)_\odot
   & = & \left( 2.6 \pm 0.6 \pm 1.4 \right) \times 10^{-5} \\
\left( \frac{\he3}{\rm H} \right)_\odot
  & = & \left( 1.5 \pm 0.2 \pm 0.3 \right) \times 10^{-5}
\eeqar
The interstellar medium (ISM) abundance for D is reported
  as \cite{linetal}
\beq
\left( \frac{\rm D}{\rm H} \right)_{\rm ISM}
  = \left( 1.6 \pm 0.09 ^{+0.05} _{- 0.10} \right) \times 10^{-5}
\eeq
however the error bar may be misleading, as recently
reported preliminary measurements of D in other lines of sight,
with abundances as low as D/H $\sim 0.5 \times 10^{-5}$ \cite{linsky}.
 Such variations
are surprising in light of chemical evolution predictions for the slow
evolution of D, and further suggest that chemical evolution predictions
for this element are to be viewed with caution.
At any rate, D chemical evolution, while model-dependent in it details,
has a clear general trend:  D decreases with time as D is only destroyed
in Galactic processing.  Thus both the solar and the ISM abundances
should be {\it lower} than the primordial abundance. The ISM \he3 abundances
 also show a large dispersion and are found in the range \he3/H $\simeq 1 - 5
\times 10^{-5}$ \cite{bbbrw}.

While chemical evolution models can in principle account for large
destruction factors for D/H over the age of  the galaxy, the relative
flatness of the \he3/H evolution is very difficult to explain.
Indeed, if we had confidence in the predictions of galactic chemical
evolution one would be able to constrain the primordial D abundance
 \cite{hata1} and ultimately the consistency of BBN
\cite{hata2}. Because stars in their
pre-main-sequence stage convert D to \he3, a high
primordial D abundance usually leads to an increasing \he3 over time
and when \he3 production in low mass stars is included the problem
becomes more acute  \cite{orstv}. In this case even
dramatic changes to standard models of chemical evolution fail,
suggesting that perhaps part of the problem lies in the stellar
evolutionary predictions for \he3  \cite{scostv}.
However as we stated at the outset, here we will not make any assumptions
regarding chemical (or stellar) evolution and use only \he4 and \li7
to test for consistency and therefore make predictions regarding
the primordial values of D and \he3.

Finally, a recent and very exciting development is the improvement of
spectral resolution in Lyman-$\alpha$ forest allows for the possibility
to observe D at high-redshift QSO absorption line systems.
   A solid D abundance
for such systems would be of the utmost interest, as these primitive
systems have not suffered much evolution (though they do contain {\it some}
metals) and so will show a D abundance much nearer to its primordial value.
Indeed, such observations have already been reported, with initial published
values being surprisingly high, with D/H $\simeq 2 \times 10^{-4}$
 \cite{quas1}.
However, with the report of a much lower abundance in a different
line of sight \cite{quas2}, the situation has
become confused.  We feel that this technique is too new to
  provide a basis for
an evaluation of BBN, but clearly this method may come to provide a strong
and clean test of BBN.

\bigskip

There is one unknown parameter in the standard model
 of big bang nucleosynthesis, the baryon to photon ratio, $\eta$.
For a given value of $\eta$, the only real uncertainty in the
calculation of the light element abundances comes from the uncertainties
in the nuclear (and weak) interaction rates employed.
Thus one can obtain a distribution of abundances at each value of
$\eta$ based on these uncertainties which we assume are Gaussian distributed.
Here, we will use\footnote{We thank Dave Thomas for his invaluable
assistance here.} the Monte Carlo results from Hata et al. \cite{hata1}.
We therefore have a likelihood distribution (unnormalized)
from the BBN calculation,
\beq
L_{\rm BBN}(Y,Y_{\rm BBN})
  = e^{-\left(Y-Y_{\rm BBN}\left(\eta\right)\right)^2/2\sigma_1^2}
\eeq
where $Y_{\rm BBN}(\eta)$ is the central value for the \he4 mass fraction
produced in the big bang, and $\sigma_1$ is the uncertainty in that
value derived from the Monte Carlo calculations.

There is also a likelihood distribution based on the observations.
In this case we have two sources of errors as discussed above, a
statistical uncertainty, $\sigma_2$ and a systematic uncertainty,
$\sigma_{\rm sys}$.  For the most part we will assume that the
systematic error is described by a top hat distribution \cite{hata1,osb}.
The convolution of the top hat distribution and the Gaussian (to describe
the statistical errors in the observations) results in the difference
of two error functions
\beq
L_{\rm O}(Y,Y_{\rm O}) =
{\rm erf}\left({Y - Y_{\rm O} + \sigma_{\rm sys}
     \over \sqrt{2} \sigma_2}\right) -
{\rm erf}\left({Y - Y_{\rm O} - \sigma_{\rm sys}
     \over \sqrt{2} \sigma_2}\right)
\label{erf}
\eeq
where in this case, $Y_{\rm O}$ is the observed
(or observationally determined)
value for the \he4 mass fraction.  As there is some doubt as to how
to treat the systematic uncertainty, we have also derived the likelihood
functions assuming that the systematic errors are Gaussian distributed.
In this case the convolution also leads to a Gaussian, with an error
$\sigma^2 = \sigma_2^2 + \sigma_{\rm sys}^2$.  Finally we have also
simply shifted the mean value $Y_{\rm O}$ by an amount
 $\pm \sigma_{\rm sys}$. In this case $L_{\rm O}$ is also a Gaussian
with spread $\sigma_2$. These functions were similarly derived for
\li7.  The asymmetric systematic errors in the top hat-Gaussian
convolution are easily incorporated: the error on the positive side
is inserted in the right error function in (\ref{erf}) while the
error on the negative side is inserted in the left error function.

For \he4 we constructed a total likelihood
function for each value of $\eta_{10} \equiv 10^{10} \eta$,
convolving for each the theoretical
and observational distributions
\beq
{L^{^4{\rm He}}}_{\rm total}(\eta) =
\int dY L_{\rm BBN}\left(Y,Y_{\rm BBN}\left(\eta\right)\right)
L_{\rm O}(Y,Y_{\rm O})
\eeq
An analogous calculation was performed for \li7.

 Of course, each observable
can individually be reconciled with the one-parameter theory.
However, when demanding that both observables be fit simultaneously,
one tests the theory.  To do this, one
examines the product of the individual likelihoods,
${L^{^4{\rm He}}}_{\rm total}(\eta) {L^{^7{\rm Li}}}_{\rm total}(\eta)$;
this gives a quantitative measure of the goodness of fit
and of the spread in the allowed values in $\eta_{10}$ (if there are any).

We first examine the case that we feel combines the most ``standard'' of
assumptions, namely:  (1) \he4 takes the central value
$\yp^0 = 0.234$ as in eq.\ (\ref{eq:yp}); (2) Li is not depleted in Pop II
stars, nor is produced it in any great quantity by cosmic-rays, i.e.,
we ignore
the second set of systematic errors $\Delta_2$;
(3) \he4 systematics are at the level
$\Delta \yp_{\rm sys} = 0.005$; (4) the systematic errors are
given by a flat (``top hat'') distribution.  With these assumptions,
we have calculated the likelihoods for \he4 and \li7; results appear
in figure \ref{fig:best-sep}.  The shapes of these curves are characteristic,
with one peak for \he4, which rises monotonically with $\eta$, and
two for \li7, which goes through a minimum.  In this case (and most others)
the minimum theoretical Li is
somewhat below most of the observational values and so the
sides of the minimum are favored, leading to the two peaks, i.e. for a
given observational value of \li7, there are two values for $\eta$ at which
this may be achieved.

The combined likelihood, for fitting both elements simultaneously,
is given by the product of the two functions in figure \ref{fig:best-sep},
and is shown  in figure \ref{fig:best-comb}.
{}From figure \ref{fig:best-sep} it is clear that \he4 overlaps
the lower \li7 peak, and so one expects that there will be concordance,
in an allowed range of $\eta$ given by the overlap region.
This is what one finds in figure \ref{fig:best-comb}, which does
show concordance, and gives an allowed (95\% CL) range of
$1.4 < \eta_{10} < 3.8$.  As we will really only be interested in the
upper limit of this range we will from here on only quote the 95\% CL
upper limit as being the upper limit of the entire range. Note that
the likelihood functions shown in figures \ref{fig:best-sep}
and \ref{fig:best-comb} are not normalized to unity.
The $\eta$ dependendant normalization has however been included.
Any further normalization would
have no effect on the predicted range for $\eta$.

Thus, for this ``standard'' case, we find that
the abundances of
\he4 and \li7 are consistent, and select an $\eta_{10}$ range which
overlaps with (at the 95\% CL) the longstanding favorite
 range around $\eta_{10} = 3$.
Further, by finding concordance  (in this case)
using only \he4 and \li7, we deduce that
if there is problem with BBN, it must arise from
D and \he3 and is thus tied to chemical evolution.
The most model-independent conclusion is that standard
BBN  with $N_\nu = 3$ is not in jeopardy,
but there may be problems with our
detailed understanding of D and particularly \he3
chemical evolution.\footnote{In fact, this conclusion
is firm only if the predicted ${\rm D}_{\rm p}$ satisfies
the only unquestioned prediction of
chemical evolution, namely that D decrease with time and so
${\rm D}_{\rm p} > {\rm D}_{\odot}$.  This is true for all
of the low--$\eta$ regions we consider.}
It is interesting to note that the central (and strongly)  peaked
value of $\eta_{10}$ determined from the combined \he4 and\li7 likelihoods
is at $\eta_{10} = 1.8$.  The corresponding value of D/H is 1.8 $\times
10^{-4}$ very close to the value  of D/H in quasar absorbers
in the published set of observations
\cite{quas1}.
It is not clear whether this is a coincidence or that we really have evidence
that three of light element abundances point to the same value of $\eta_{10}$.

When we vary some of the values or assumptions concerning systematic errors,
our 95\% CL range for $\eta$ is somewhat affected.  These results are
summarized in the table below.  Had we used the central value of
\he4 as determined by Olive and Steigman \cite{osa}, $Y_p = 0.232$
(all other assumptions held fixed) our 95\% CL upper limit shifts down
to $\eta_{10} < 3.3$. At
this point one should
take note at the sensitivity of the upper limit on $\eta$ to the  \he4
abundance. As the size of the assumed systematic error
for \he4 is sometimes questioned \cite{sg,csta} we have run our likelihood test
for $Y_p = 0.234 \pm 0.003 \pm 0.010$, i.e., we have doubled the assumed
systematic error (still treated as a top hat).  In this case there is a broad
overlap between the likelihood functions for \he4 and and both peaks for
\li7. There are
now two peaks in the product of the distributions at $\eta_{10} = 1.8$
and $\eta_{10} = 3.6$ the
95\% CL upper limit increases to $\eta_{10} < 4.5$.

\begin{table}[htb]
\label{tab:etalim}
\caption{Limits on $\eta$}
\begin{center}
\begin{tabular}{cccccc}
\hline\hline
$Y_p$ & $\Delta Y_{\rm sys}$ & sys type&
 \li7$_{\rm sys}$& $\eta_{10}^{\rm min}$ & $\eta_{10}^{\rm max}$ \\
\hline
$.234 \pm .003$ & .005& top hat & $\Delta_1$ &1.4 & 3.8 \\
\hline
 $.232 \pm .003$& .005& top hat& $\Delta_1$& 1.3 & 3.3 \\
 $.234 \pm .003$& .010& top hat& $\Delta_1$ &
1.4 & 4.5 \\
$.234 \pm .003$ & .005 &  Gaussian & $\Delta_1$& 1.3  & 4.4 \\
$.239 \pm .003$ & .000 &  shift & $\Delta_1$& 1.7 & 4.4 \\
 $.234 \pm .003$& .010& top hat& $\Delta_1 + \Delta_2$ &
1.2 & 3.8 \\
\hline\hline
\end{tabular}
\end{center}
\end{table}

If we return to our standard values and treat the systematic errors
as if they were Gaussian distributed,  then there is again  a broad
overlap between the likelihood functions for \he4 and \li7 though the two
peaks (at $\eta_{10} = 1.8$
and $\eta_{10} = 3.3$) in the product distribution overlap. The 95\% CL
upper limit to $\eta$ is now $\eta_{10} < 4.4$.  One can also imagine
the systematic errors amounting to
a shift to the central value by $\sigma_{\rm sys}$.
For \he4, we apply this shift upwards and take $Y_p = 0.239 \pm
0.003$.  For \li7, our results are sensitive to the direction we shift
the central value.  When a shift by $4 \times 10^{-11}$ is added to the
\li7 central value (making it \li7/H $= 2.02 \times 10^{-10}$)
the \li7 peaks are split farther apart as this value of \li7/H is
obtained in BBN calculations at either higher or lower values of
$\eta$.  There is now only a minimal overlap between the likelihood functions
of \he4 (whose peak now sits between the two \li7 peaks) and \li7.
The product of the likelihood functions now shows two separate peaks, however
because of the poor agreement between the two elements we discard this
possibility and do not include it in the table.
In fact, this disagreement can be quantified by taking
the product of the normalized likelihood functions. The (very) low value
of the product relative to the other cases we consider
 would be such a signal.
 If instead we apply a downward
shift of  $3 \times 10^{-11}$ in \li7, there is substantially more overlap
and the 95\% CL upper limit to $\eta$ is 3.9. When no shift is applied to
\li7, there is still a reasonable amount of overlap and there are still two
peaks in the product distribution at $\eta_{10} = 2.0$ an 3.5,
and the upper limit is
now $\eta_{10} < 4.4 $. It is this case that appears in the table.

Finally, we consider the possibility that the errors in \li7 are in fact
larger than assumed above, due to either stellar depletion or cosmic-ray
production of \li7. In this case, as the systematic errors are treated
as top hat distributions, the asymmetric uncertainties in \li7
effectively shift upward the central value
(cf. eq.(\ref{erf})). This
causes the two peaks in the \li7 likelihood distribution to move apart,
as we have seen above.
The \he4 likelihood distributions is now almost entirely under the low
$\eta$ peak of the \li7 distribution. The result is an upper limit,
$\eta_{10} < 3.8$. The peak of the product
 distribution is at $\eta_{10} = 1.7$. In general we found that, contrary to
 the na\"{\i}ve expectation,
when the \li7 uncertainties increase substantially (when the effects of
depletion are allowed for) the range for BBN consistency actually
shrinks rather than expands.

Having found the allowed range of $\eta$, we now turn to the
predictions for primordial D and \he3.
Since  D and \he3 are monotonic functions of $\eta$, a prediction for
$\eta$, based on \he4 and \li7, can be turned into a prediction for
D and \he3.  In figure \ref{fig:dhe}, we show the abundances of
D and \he3 as a function of $\eta_{10}$ along with the one $\sigma$
uncertainty in the calculations from the Monte Carlo results of \cite{hata1}.
We also show by a set of rectangles the 68\% (dashed)
and 95\% CL (dotted) ranges for
D and \he3 as given by our likelihood analysis above. The corresponding 95\%
CL ranges are D/H  $= (5.5 - 27)  \times 10^{-5}$ and
and \he3/H $= (1.4 - 2.7)  \times 10^{-5}$.
Again, any potential inconsistency with BBN must be related to
D and \he3 and thus from our
more detailed understanding (or lack thereof) of
chemical (or stellar) evolution.

We would like to stress that in essentially all of the cases we considered
(with the possible exception of shifting up the observed
\he4 abundance while at the same time shifting up the \li7
abundance) we found a broad region of concordance between
the predictions of BBN and the observations of \he4 and \li7.
This region of concordance in what we deemed the most standard case
allowed values of $\eta$ are as high as 3.8 at the 95\% CL. We note
that this region overlaps with the one found in \cite{hata2,cstb}
but that the overlap is in the ``2 $\sigma$" error bars on each side.
More importantly, this upper limit as well as
 lower values of $\eta$ can easily
accommodate the evolution of D \cite{vop,orstv}. The main problem
lies with \he3. Indeed, it was argued in \cite{hata2} that a
high value of $\eta$ corresponding to a relatively low value of
D/H was necessary to account for the evolution of \he3.
However, when the effects of \he3 production in low mass
stars is included \cite{orstv,galli,dst} even the most successful
models of galactic chemical evolution fail to explain the
observed abundances of \he3.  Thus, as these models are not
capable of explaining the \he3 abundances at high $\eta$, they can
not be used as a constraint on BBN forbidding lower values of $\eta$.
We believe that the values derived here must ultimately be
incorporated in any model of chemical evolution.

Of all the light element abundances, \he4 and \li7 are the least
dependent on specific models of galactic chemical evolution.
In using the best observationally determined values for these elements
as well as simple assumptions concerning the treatment of systematic errors,
we have found concordance in the one parameter theory of standard big
bang nucleosynthesis with $N_\nu = 3$. Though our prediction for
the primordial value for D/H is consistent with chemical evolution models
we know of no models at present which can account for the evolution
of \he3 as implied by the observations of \he3.  We also know of no ``standard"
model which can account for the evolution of \he3 at higher values
of $\eta$ (lower values of D/H) when the production of \he3 in
 low mass stars is included.  Of course, if a firm value for primordial
D/H can be established from the observations of quasar absorption
systems, these predictions will tested.

\bigskip

We would like to thank Craig Copi, David Schramm, Gary Steigman, and
for useful discussions, and we are especially grateful to Dave Thomas
for his help with the calculations as well as helpful advice.
This material is based upon work supported by the North Atlantic Treaty
Organization under a Grant awarded in 1994.
This work was supported in part by
DOE grant DE-FG02-94ER40823.

\newpage

\centerline{Figure Captions}

\bigskip

\begin{enumerate}

\item \label{fig:best-sep}
Likelihood distribution for each of \he4 and \li7, shown as a
function of $\eta$.  The one-peak structure of the \he4 curve
corresponds to its monotonic increase with $\eta$, while
the two-peaks for \li7 arise from its passing through a minimum.

\item \label{fig:best-comb}
Combined likelihood for simultaneously fitting \he4 and \li7,
as a function of $\eta$.

\item  \label{fig:dhe} D/H and \he3/H as a function of $\eta_{10}$
from BBN along with the one $\sigma$ uncertainty from
Monte Carlo calculations \cite{hata1}. Also shown are the values (demarcated
by rectangles) of
D/H and \he3/H consistent with
68\% (dashed) and 95\% CL (dotted)  likelihood values for $\eta_{10}$.

\end{enumerate}

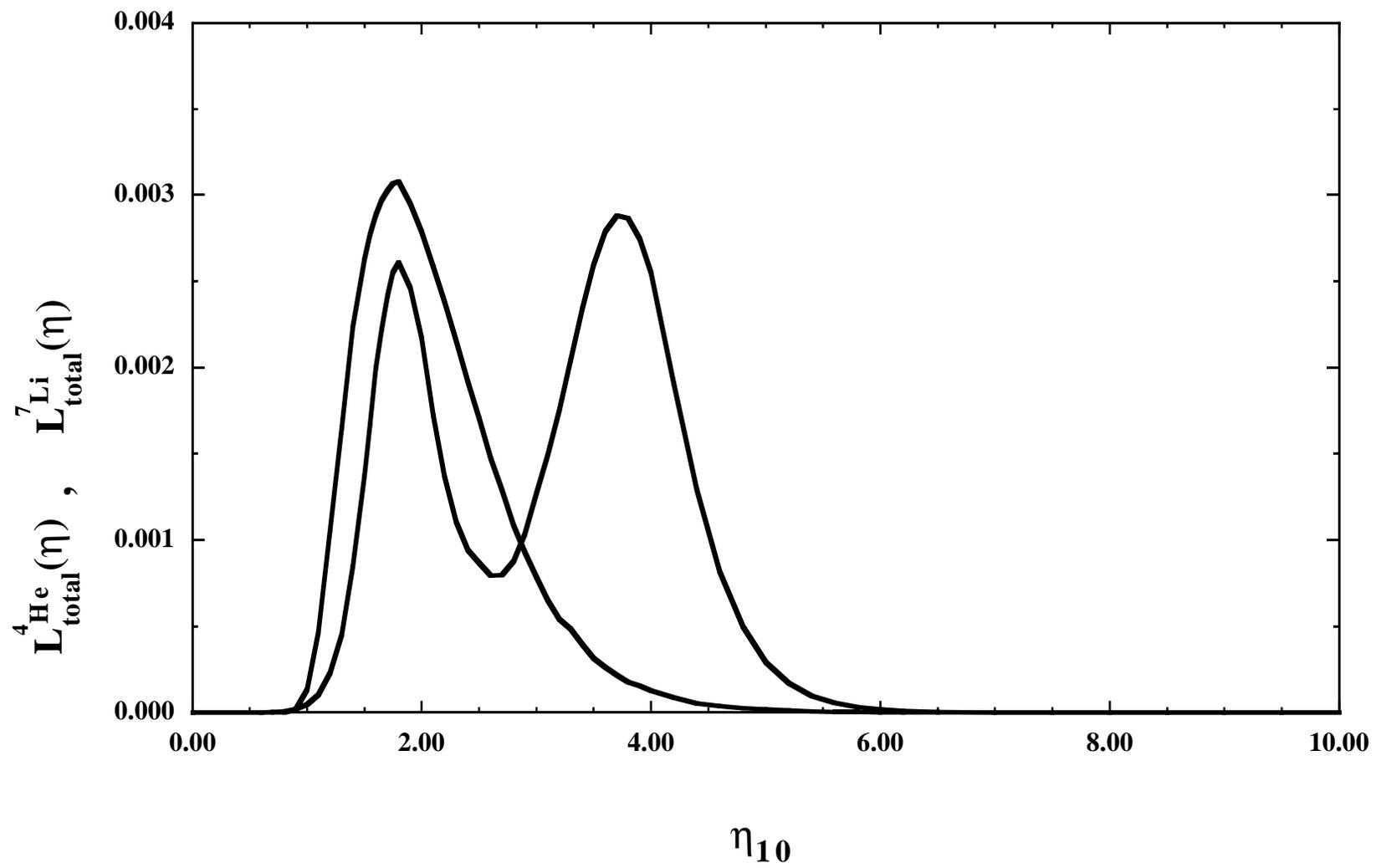
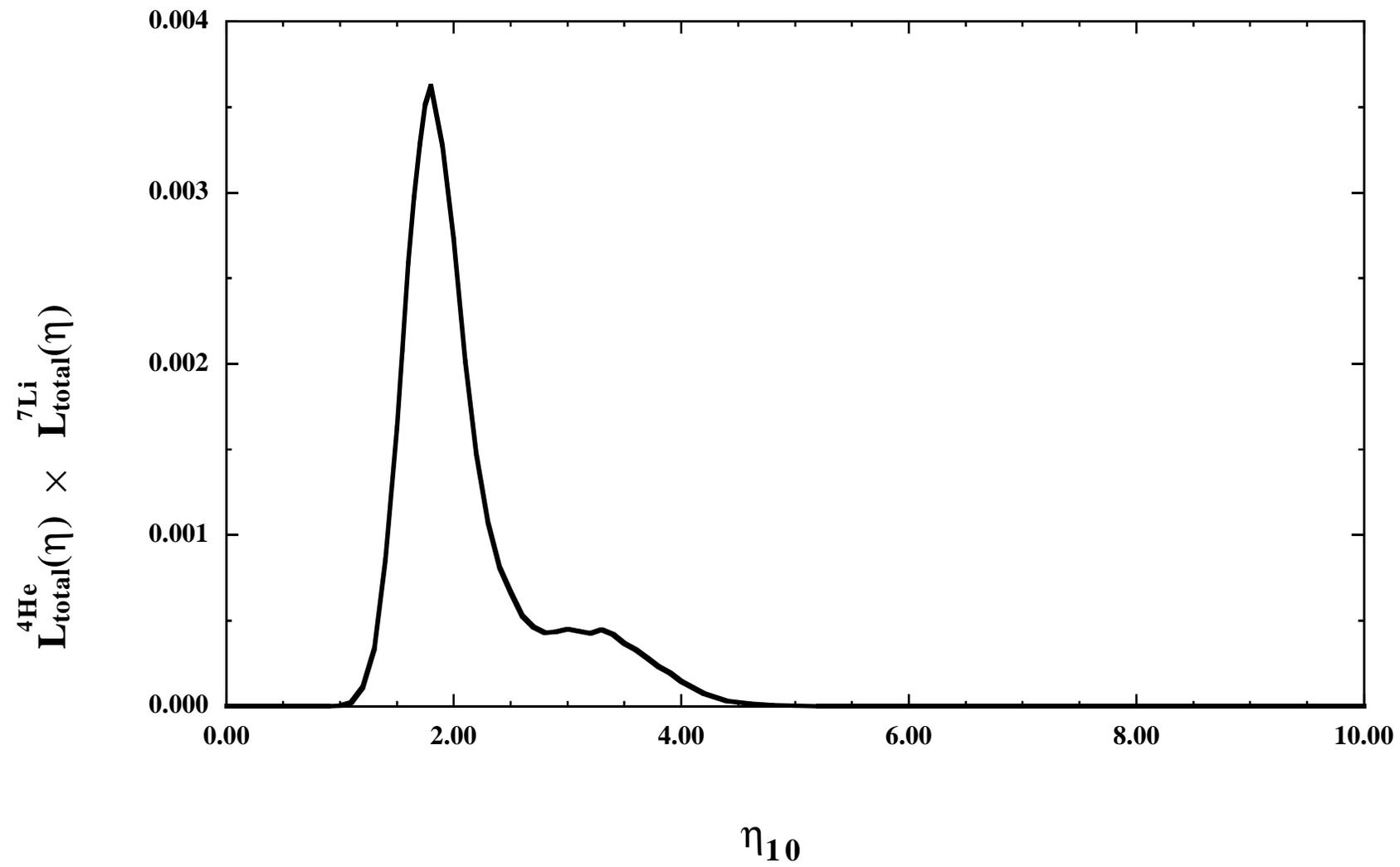
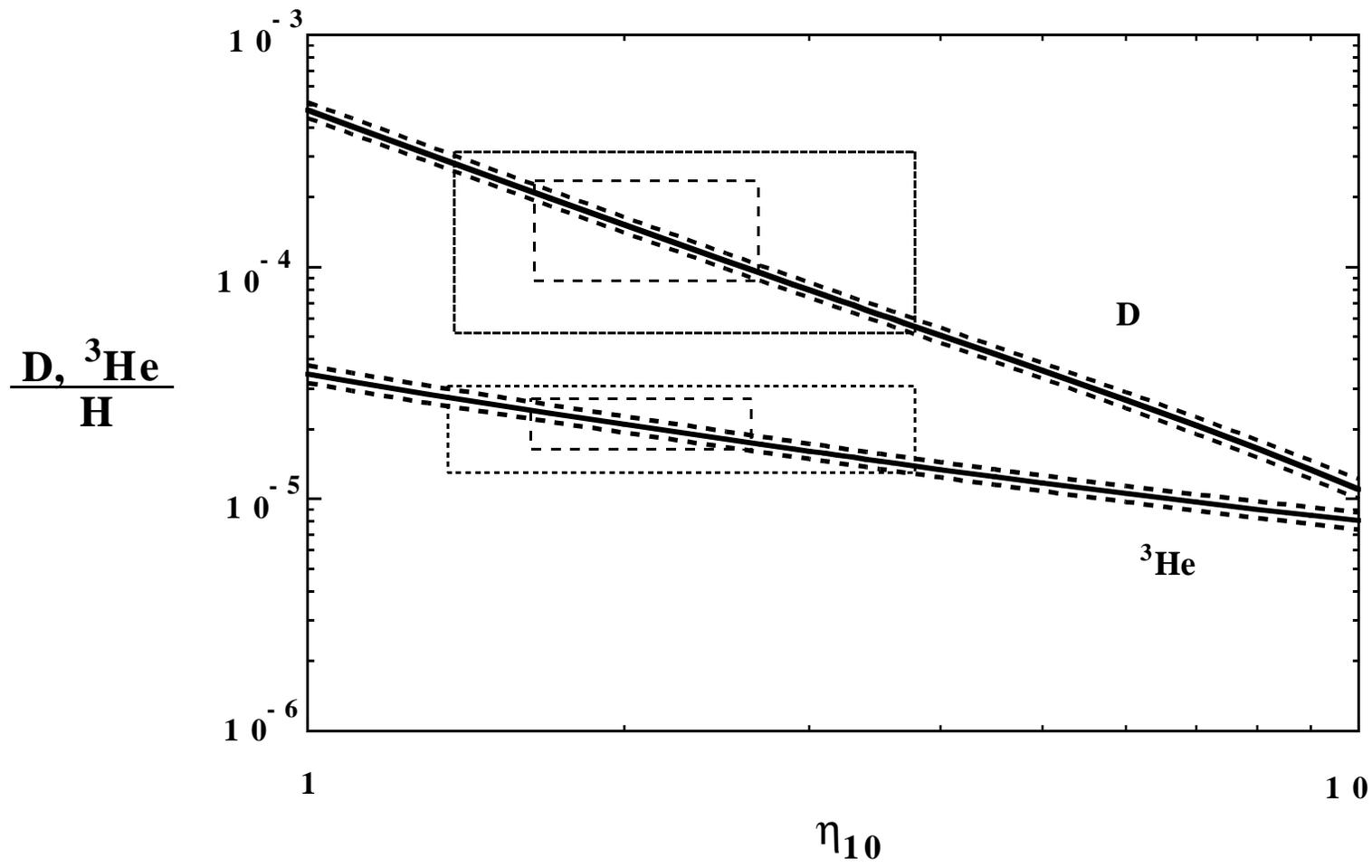
\end{document}